\documentclass[aps,prb,twocolumn]{revtex4}

\usepackage{chemformula} % Formula subscripts using \ch{}
\usepackage[T1]{fontenc} % Use modern font encodings

\usepackage{amsmath,amsfonts,bm}
\usepackage{graphicx}  %epsfig,
\usepackage{color}
\usepackage{extarrows}
\usepackage{enumitem}
\usepackage{empheq}

\renewcommand{\d}{{\rm d}}

\newcommand{\ud}{{\mathrm{d}}}

\newcommand{\w}{\omega}

\newcommand{\ti}{\tilde}
\newcommand{\A}{\mbox{\tiny A}}
\newcommand{\B}{\mbox{\tiny B}}
\newcommand{\D}{\mbox{\tiny D}}

\newcommand{\E}{\mbox{\tiny E}}
\newcommand{\F}{\mbox{\tiny F}}

\newcommand{\tS}{\mbox{\tiny S}}
\newcommand{\T}{\mbox{\tiny T}}

\newcommand{\dg}{\dagger}
\newcommand{\la}{\langle}
\newcommand{\ra}{\rangle}

\newcommand{\ET}{\mbox{\tiny ET}}

\newcommand{\nl}{\nonumber \\}
\newcommand{\be}{\begin{equation}}
\newcommand{\ee}{\end{equation}}
\newcommand{\bsube}{\begin{subequations}}
\newcommand{\esube}{\end{subequations}}
\newcommand{\Eq}[1]{Eq.\,(\ref{#1})}

\newcommand{\Fig}[1]{Fig.\,\ref{#1}}

\newcommand{\RN}[1]{%
  \textup{\uppercase\expandafter{\romannumeral#1}}%
}

\allowdisplaybreaks[1]

\begin{document}

\title{Electron transfer under the Floquet modulation in donor--bridge--acceptor systems}

\author{Yu Su}
\author{Zi-Hao Chen}
\author{Haojie Zhu}
\author{Yao Wang} 
\email{wy2010@ustc.edu.cn}
\author{Lu Han}
\author{Rui-Xue Xu}
%\email{rxxu@ustc.edu.cn}
\author{YiJing Yan}
\email{yanyj@ustc.edu.cn}
\affiliation{%Hefei National Laboratory for Physical Sciences at the Microscale and
 Department of Chemical Physics,
University of Science and Technology of China, Hefei, Anhui 230026, China}

%\abbreviations{IR,NMR,UV}
% \keywords{Electron transfer, Floquet modulation, Donor--bridge--acceptor system, Rate kernel}

% \begin{tocentry}

%   \begin{center}
%     \includegraphics[width=1\columnwidth]{./fig1.eps}
%   \end{center}
% \end{tocentry}

\begin{abstract}
  Electron transfer (ET) processes are of broad interest in modern chemistry.
  With the advancements of experimental techniques, one may modulate the ET via such as the light--matter interactions.
  In this work, we study the ET under a Floquet modulation occurring in the donor--bridge--acceptor systems, with the rate kernels projected out from the exact disspaton equation of motion formalism.
  %The exact rate kernels are constructed and evaluated via the projected disspaton--equation--of--motion approach.
  %
  This together with the Floquet theorem enables us to investigate the interplay between the intrinsic non-Markovianity and the driving periodicity.
  The observed rate kernel exhibits a Herzberg--Teller--like mechanism induced by the bridge fluctuation subject to effective modulation.
\end{abstract}

% \clearpage
\maketitle

% p1
\paragraph*{Introduction.} Electron transfer (ET) processes are of broad interest in modern chemistry, \cite{Mar56966,Mar64155,Sum864894,Mar93599} 
and many of them occur in the donor--bridge--acceptor (DBA) scenarios. \cite{Del141492,Bix9935,New01,Sko01,Nit01681,Tro035782,Gal07103201,Zha09034111}
The bridge, which remains itself before and after the reaction, could be considered as a rigid spacer within an intramolecular ET system. \cite{Tro035782}
In condensed phases, the solvent environment also plays a crucial role. \cite{Wol871957,Spa873938,Spa883263,Spa884300,Han0611438}
Fluctuations of both the bridge and the solvent will manifestly affect the rate of ET processes.
From a theoretical point of view, one can in principle exactly construct the generalized rate equation,
\begin{align}\label{gre}
  \dot P_{\D}(t)= -\int^{t}_{0}\!\d\tau\, k(t-\tau;t)P_{\D}(\tau)+\int^{t}_{0}\!\d\tau\, k'(t-\tau;t)P_{\A}(\tau).
\end{align}
Here, $P_{\D}(t)$ and $P_{\A}(t)$ are the donor and acceptor populations, respectively.
The forward and backward rate memory kernels, $k(\tau;t)$ and $k'(\tau;t)$, have involved the influences of the bridge fluctuations, the solvent effects and the possibly existent external modulations.\cite{Gon15084103,Zha163241}
Here, the variable $\tau$ characterizes the memory timescale, i.e., the non-Markovianity, and $t$ represents the time--dependence due to the external fields.

 With the advancements of experimental techniques, one can modulate the ET process via such as light--matter interactions. 
 Especially, if the external modulation is periodic, it is called the Floquet modulation.\cite{Buk15139, Mik16144307, Res16250401,Tha181243,Eng21090601}
In this Letter, we investigate how the ET rate kernels will be influenced by the fluctuating bridges and Floquet modulation, with attention to the intrinsic non-Markovianity and the external periodicity.
 Specifically, we will focus on the case in which the energy difference between the donor state $|\rm D\ra$ and the acceptor state $|\rm A\ra$ is periodically modulated, which can be realized via the field--dipole interaction or the Stark effect. \cite{Tha181243} 
We first give a perturbative analysis, followed by the discussion on the influences of the fluctuating bridge and the period of modulation.
  Then the non-Markovian rate kernels in \Eq{gre} are projected out from the exact dissipaton equation of motion (DEOM).\cite{Yan14054105} 
 This is the second quantization generation of the notable hierarchical equations of motion (HEOM) formalism,
\cite{Tan906676,Tan06082001,Yan04216,
Xu05041103,Xu07031107,Jin08234703} covering both the reduced system and hybrid bath modes dynamics. \cite{Yan14054105, Zha15024112, Zha18780, Wan20041102}
The linear space algebra of DEOM facilitates the utilization of Nakajima--Zwanzig projection operator technique, so that we can focus on any subspace dynamics and construct non-Markovian rate kernels. \cite{Zha163241}
 The rate kernels are investigated with a DBA model system via both numerical and analytical methods in the DEOM framework. Especially, we pay attentions to the interplay between the intrinsic non-Markovianity and the driving periodicity, with the help of Floquet theorem.
The observed rate kernel exhibits a Herzberg--Teller--like mechanism induced by the bridge fluctuation subject to effective modulation.
% We pay special attentions to the interplay between the intrinsic non-Markovianity and the driving periodicity, and it is discussed with the help of the Floquet theorem.
 %
Throughout this Letter, we set $\hbar=1$ and $\beta=1/(k_BT)$ with $k_{B}$ being the Boltzmann constant
and $T$ the temperature.

\begin{figure}[h]
  \includegraphics[width=0.85\columnwidth]{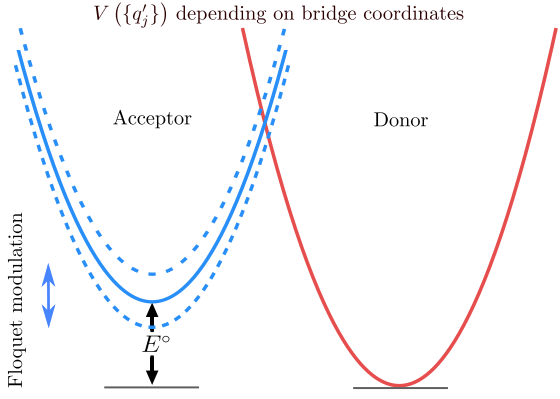}
    \caption{Schematics of ET under the Floquet modulation in the DBA system.
  }
  \label{fig1}
  \end{figure}

\paragraph*{Theoretical model and perturbative analysis.} Consider an ET DBA system with the total composite Hamiltonian,
% \bsube
\begin{align}\label{Het}
  H_{\ET}&=h_{\D}|{\rm D}\ra\la {\rm D}|+(E^{\circ}+h_{\A})|{\rm A}\ra\la {\rm A}|+H_{\B} \nl
  &\quad +V\big(\{\ti q_k\}\big)\left(|{\rm D}\ra\la {\rm A}|+|{\rm A}\ra\la {\rm D}|\right ).
\end{align}
% Here, the bridge--dependent nonadiabatic transfer coupling term $\hat V$ 
% \esube
Here $V\big(\{\ti q_k\}\big)$ depends on the bridge coordinates and the fluctuating bridges Hamiltonian is 
 $
    H_{\B}=\sum_k \frac{\ti \omega_k}{2} \left(\ti p_k^{2}+\ti q_k^{2}\right)$.
In \Eq{Het}, $E^{\circ} \simeq \Delta_{\rm r} G^{\circ}$ amounts to the standard reaction Gibbs energy, for the electron transferring from $|{\rm D}\ra$ to $|{\rm A}\ra$. 
Donor and acceptor are both associated with their own solvent environments, $h_{\D}$ and $h_{\A}$, 
being $h_{\D}=\sum_j \frac{\omega_j}{2}\left(p_j^2+x_j^2\right)$ and $h_{\A}=\sum_j \frac{\omega_j}{2} \left[p_j^2+(x_j-d_{j})^2\right]$, respectively.
%Here, the modes of solvent and bridge are independent and will be treated separately. 
The total ET composite was initially $\rho_{\T}(t_0)=\rho^{{\rm eq}}(T)\otimes|{\rm D}\ra\la {\rm D}|$, the thermal equilibrium in the donor state, with $\rho^{{\rm eq}}(T)\equiv (e^{-\beta h_{\D}}/{\rm tr}e^{-\beta h_{\D}})\otimes (e^{-\beta H_{\B}}/{\rm tr}e^{-\beta H_{\B}})$.  
 Floquet modulation leads to the periodic changes of $E^{\circ}$,  as
\begin{align}
  E^{\circ} \longrightarrow E(t)=E^{\circ}+{\cal E}\cos(\Omega t)
\end{align}
with the amplitude of ${\cal E}$ and the frequency of $\Omega$. 
We sketch these theoretical settings in \Fig{fig1}.
  
In the absence of modulation (${\cal E}=0$), the standard perturbation theory gives the forward rate constant the expression:
  \cite{Tro035782,Zha09034111}
\bsube
\begin{align}\label{sss}
  K_{0}=2{\rm Re}\int_{0}^{\infty}\!\!{\rm d}t\,\,v(t)e^{-i(E^{\circ}+\lambda)t}e^{-g(t)},
\end{align}
where $v(t)=\big\la V(\{\ti q_k(t)\})V(\{\ti q_k\})\big\ra_{\B}$ with $\ti q_k(t)\equiv e^{iH_{\B}t}\ti q_k e^{-iH_{\B}t}$ and $\la\,\cdot\, \ra_{\B}\equiv{\rm tr} (\,\cdot\,e^{-\beta H_{\B}})/{\rm tr}e^{-\beta H_{\B}}$. 
In \Eq{sss}, $\lambda\equiv \la \hat  U\ra_{\D}\equiv \la h_{\A}-h_{\D}\ra_{\D}$ and 
\be 
  g(t)=\int_{0}^{t}\!{\rm d}\tau\int_{0}^{\tau}\!{\rm d}\tau'\, C(\tau-\tau'),
\ee
\esube
where the correlation $ C(t)\equiv \la \delta \hat U(t)\delta \hat U\ra_{\D}
$ with $\delta \hat U=\hat U-\lambda$ and 
 $\delta \hat  U(t)\equiv e^{ih_{\D}t}\delta \hat  Ue^{-ih_{\D}t}$. 

In the presence of modulation (${\cal E}\neq 0$), the time--dependent Hamiltonian  $H_{\T}(t)$ can be recast as
\bsube\label{5}
\begin{align}\label{eq10}
  H_{\T}(t) = H\left(t;\{\ti q_k\}\right)+h_{\D}+h_{\B} - |{\rm A}\ra\la {\rm A}|\delta \hat U
\end{align}
with
\begin{align}\label{eq11}
  H(t;\{\ti q_k\}) &= (E^\circ + {\cal E}\cos\Omega t+\lambda)|{\rm A}\ra\la {\rm A}| + \hat V
\end{align}
and 
\begin{align}\label{eq11b}
  \hat V = V(\{\ti q_k\})\left(|{\rm D}\ra\la {\rm A}|+|{\rm A}\ra\la {\rm D}|\right).
\end{align}
\esube
%with reorganization energy $\lambda\equiv \sum_j\omega_jd_j^2/2$.
To proceed, one may employ a unitary transformation generated by
\begin{align}
  \Lambda(t)\equiv e^{-i\varphi(t)|{\rm A}\ra\la {\rm A}|}\ \  \text{with}\ \  \varphi(t) = {\cal E}\sin(\Omega t)/\Omega.
\end{align}
Under such a unitary transformation, we obtain the new Hamiltonian 
\bsube\label{7}
\begin{align}\label{eq12}
  H_{\T}'(t)& = \Lambda^\dagger(t) H_{\T}(t)\Lambda(t) - i\Lambda^\dagger(t)\frac{\partial}{\partial t}\Lambda(t)
  \nl &
  = H'\left(t;\{\ti q_k\}\right) + h_{\D}+H_{\B} - |{\rm A}\ra\la {\rm A}|\delta U,
\end{align}
where
\begin{align}\label{eq13}
  H'(t;\{\ti q_k\})=(E^{\circ}+\lambda)|{\rm A}\ra\la {\rm A}| +\hat V'(t) 
\end{align}
with
\begin{align}\label{eq14}
  \hat V'(t) =V(\{\ti q_k\})\left(e^{-i\varphi(t)}|{\rm D}\ra\la {\rm A}|+e^{i\varphi(t)}|{\rm A}\ra\la {\rm D}|\right).
\end{align}
\esube
It is easy to check, \Eq{5} generates the same population dynamics as \Eq{7}.
In the high-frequency limit, we could take the time average of $H_{\T}'(t)$ over one period $T_0\equiv 2\pi/ \Omega$ to modify the rate constant given by perturbation theory [\rm cf.\,\Eq{sss}]. To this end, we do the approximation
$
  {H}'_{\T}(t)\longrightarrow \tilde{H}_{\T} \equiv \frac{1}{T_0}\int_0^{T_0}\!\ud t\, H_{\T}'(t), 
$
and this amounts to 
$
\hat V(t)\rightarrow \tilde{V} \equiv V(\{\ti q_k\})J_0({\cal E}/\Omega)
$
with $J_0(z)$ being the zeroth order Bessel function.
Therefore, the nonadiabatic rate under the high--frequency Floquet modulation reads
\begin{align}\label{k_ref}
  K_{0} = 2J_0^2({\cal E}/\Omega){\rm Re}\!\int_{0}^{\infty}\!\!{\rm d}t\,\upsilon(t)e^{-i(E^{\circ}+\lambda)t}e^{-g(t)}.
\end{align}
This perturbative rate formula serves as a reference for the following nonperturbative exhibitions.

\paragraph*{Projected DEOM to rate kernels.} For illustrations, we assume $V(\{\ti q_k\})$ the form of \cite{Zha09034111}
\begin{align}
  V(\{\ti q_k\})\equiv \la V\ra_{\B}-\delta \hat V =  \la V\ra_{\B} - \sum_k \ti c_k \ti q_k.
\end{align}
This is the scenarios that can be exactly handled by DEOM--space quantum mechanics. To proceed we introduce
\begin{align}
  v(t)=\la V\ra_{\B}^2+\la \delta \hat V(t)\delta \hat V\ra_{\B}=\la V\ra_{\B}^2+\sum_j{\ti c^{2}_j}\la \ti q_{j}(t)\ti q_{j}\ra_{\B}.
\end{align}
 Now we can construct the DEOM based on  \Eq{5}, or equivalently \Eq{7}:
(i) In the former case, the system--plus--environment decomposition reads
\begin{align}\label{HSB1}
  H_{\T}(t) = H_{\tS}(t)+h_{\E} - |{\rm A}\ra\la {\rm A}|\delta \hat U-\hat Q\delta \hat V,
\end{align}
with $h_{\E}=h_{\D}+H_{\B}$, $H_{\tS}=(E^\circ + {\cal E}\cos\Omega t+\lambda)|{\rm A}\ra\la {\rm A}| + \la V\ra_{\B}\left(|{\rm D}\ra\la {\rm A}|+|{\rm A}\ra\la {\rm D}|\right )$, and $\hat Q=|{\rm D}\ra\la {\rm A}|+|{\rm A}\ra\la {\rm D}|$. In this case, the system Hamiltonian is time--dependent, while the dissipative mode $\hat Q$ is not;
(ii) In the latter case, 
\begin{align}\label{HSB2}
  H_{\T}'(t) = H'_{\tS}(t)+h_{\E} - |{\rm A}\ra\la {\rm A}|\delta \hat U-\hat Q'(t)\delta \hat V.
\end{align}
Compared with \Eq{HSB1}, the $H'_{\tS}(t)=(E^{\circ}+\lambda)|{\rm A}\ra\la {\rm A}|+\la V\ra_{\B}\left(e^{-i\varphi(t)}|{\rm D}\ra\la {\rm A}|+e^{i\varphi(t)}|{\rm A}\ra\la {\rm D}|\right) $ and $\hat Q'(t)=\left(e^{-i\varphi(t)}|{\rm D}\ra\la {\rm A}|+e^{i\varphi(t)}|{\rm A}\ra\la {\rm D}|\right)$ are different from the former case. Both the system Hamiltonian and the dissipative mode $\hat Q'(t)$ are time--dependent.

The rate kernels constructed from \Eq{HSB1} should be exactly the same with that  from \Eq{HSB2}, and our numerical results validate this point. 
The rate kernels are constructed via the DEOM approach. 
Based on the composite Hamiltonian in \Eq{HSB1} or \Eq{HSB2}, we can write the DEOM  in the form of
\be\label{sdeom}
\dot{\bm\rho}(t)=-i\bm{\mathcal{L}}(t){\bm\rho}(t).
\ee
This resembles $\dot{\rho}_{\T}=-i{\cal L}_{\T}(t)\rho_{\T}$ with mapping the total system--plus--bath composite  Liouvillian to  the DEOM--space dynamics generator, ${\cal L}_{\T}(t)\rightarrow \bm{\mathcal{L}}(t)$,
and
$
\rho_{\T}(t)\rightarrow {\bm \rho}(t)=\{\rho_{\bf n}^{(n)}(t); n=0,1,2,\cdots\}.
$ 
Here, ${\cal L}_{\T}(t)\equiv [H_{\T}(t),\,\cdot\,]$ in case (i) or  $[H'_{\T}(t),\,\cdot\,]$ in case (ii). 
We will leave the detailed  information of the DEOM (\ref{sdeom}) in Supplementary material (SM).
%
%From the DEOM (\ref{sdeom}), we know that the propagator
%\be
%  \bm{{\cal G}}(t,\tau)=\exp_+\Big[-i\int_{\tau}^{t}\d\tau'\,\bm{{\cal L}}(\tau')\Big]
%\ee
%governs the dynamics of the DEOM state function $\bm{\rho}\equiv\{\rho_{\bf n}^{(n)}\}$.
%
% 
To proceed, define DEOM--space projection operators, $\bm{{\cal P}}$ and $\bm{{\cal Q}}=\bm{{\cal I}}-\bm{{\cal P}}$, for partitioning $\bm{\rho}\equiv\{\rho_{\bf n}^{(n)}\}$ into the population and coherence components,
respectively:\cite{Zha163241}
\be\label{HEOM_PQ_def}
\begin{split}
 \bm{{\cal P}}\bm{\rho}(t)&= \Big\{\sum_{a} \rho^{(0)}_{aa}(t)|a\ra\la a|;\ 0,\,0,\,\cdots\,\Big\} \equiv {\bm p}(t)\,,
\\
 \bm{{\cal Q}}\bm{\rho}(t)&= \Big\{\sum_{a\neq b} \rho^{(0)}_{ab}(t)|a\ra\la b|;\ \rho^{(n>0)}_{\bf n}(t)\Big\} \equiv {\bm\sigma}(t)\,.
\end{split}
\ee
We can now recast the DEOM (\ref{sdeom}) in terms of
\be\label{sep}
\begin{bmatrix}
	\dot{\bm p}(t) \\ \dot{\bm\sigma}(t)
\end{bmatrix}
=-i \begin{bmatrix}
 	\bm{{\cal P}{\cal L}}(t)\bm{{\cal P}}  & \bm{{\cal P}{\cal L}}(t)\bm{{\cal Q}}\\
	\bm{{\cal Q}{\cal L}}(t)\bm{{\cal P}}  & \bm{{\cal Q}{\cal L}}(t)\bm{{\cal Q}}
\end{bmatrix}
\begin{bmatrix}
	{\bm p}(t) \\ {\bm\sigma}(t)
\end{bmatrix}.
\ee
After some simple algebra we obtain \cite{Zha163241}
\be\label{popt}
  \dot{\bm p}(t)= \int_{0}^{t}\!\d\tau\,\bm{\tilde{K}}(t-\tau;t){\bm p}(\tau),
\ee
with the rate kernel being formally of
\bsube
\be\label{calK_formal}
 \bm{\tilde{K}}(t-\tau;t) = -\bm{{\cal PL}}(t)\bm{{\cal Q}}
 \bm{{\cal U}}(t,\tau)
\bm{{\cal QL}}(\tau)\bm{{\cal P}},
\ee
with
\be
  \bm{{\cal U}}(t,\tau)\equiv \exp_{+}\Big[-i\!\int_{\tau}^{t}\!\d\tau'\,\bm{{\cal L}}(\tau')\Big].
\ee
\esube
Apparently, $-k(t-\tau;t)$ and $k'(t-\tau;t)$ in \Eq{gre}  are the $|{\rm D}\ra\la {\rm D}| \rightarrow |{\rm D}\ra\la{\rm D} |$ and $|{\rm A}\ra\la {\rm A}| \rightarrow |{\rm D}\ra\la{\rm D} |$ components of $\bm{\tilde{K}}(t-\tau;t)$, respectively. 

%That is to say,
%\begin{align}
%-k(t-\tau;t)&=\big\la\!\big\la  |{\rm D}\ra\la{\rm D} |, 0,0, \cdots \big|\big| \bm{\tilde{K}}(t-\tau;t)\big|\big|\,|{\rm D}\ra\la{\rm D} |, 0,0, \cdots\big\ra\!\big\ra ,
%\\
%k'(t-\tau;t)&=\big\la\!\big\la  |{\rm D}\ra\la{\rm D} |, 0,0, \cdots \big|\big| \bm{\tilde{K}}(t-\tau;t)\big|\big|\,|{\rm A}\ra\la{\rm A} |, 0,0, \cdots\big\ra\!\big\ra.
%\end{align}

\paragraph*{Rate kernel analysis.} In the following, we explicitly illustrate some key properties of the forward rate kernel $k(\tau;t)$ in \Eq{gre}, with the help of numerical examples. The analysis on $k'(\tau;t)$ is similar and thus omitted due to the limitation of space.
 It is worth noting the periodicity, 
% \begin{align}
  $k(\tau;t) = k(\tau;t+T_0)$.
% \end{align}
The rigorous proofs are to be found in SM. 
\begin{figure}
  \includegraphics[width=1\columnwidth]{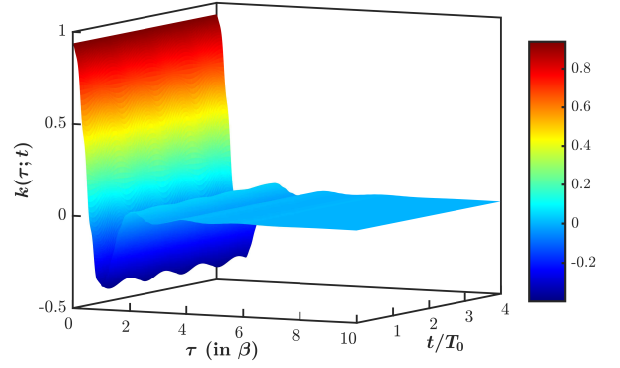}
    \caption{An example of rate kernel $k(\tau;t)$, in unit of $\beta^{-2}$. We adopt ${\cal E}=2$, $E^{\circ}=1.5$ and  $\la V\ra_{\B}=0.2$ [cf.\,\Eq{HSB1}]. Besides, $\lambda=\lambda'=0.2$ and $\gamma=\w_0=\zeta=1$ [cf.\,\Eq{J1}]. All these parameters are in units of $\beta^{-1}$.
  }
  \label{fig2}
\end{figure}
In \Fig{fig2}, we explicitly exhibit an example of the computed rate kernels. The kernel $k(\tau;t)$ is plotted with respect to  $\tau$ and $t$. The external modulation frequency adopts $\beta\Omega=8\pi$. 
As shown in \Fig{fig2}, the kernel is periodic ($T_0=2\pi/\Omega$) with respect to $t$, and damping along the memory length, $\tau$. 
%For comparison, we also plot a  relatively low--frequency case, $\beta\Omega=0.5$, in the right panel. 
In the simulations, we model the spectral densities,  
$J_{\D}(\w)\equiv\frac{1}{2}\int^{\infty}_{-\infty}\!\d t\,
  e^{i\w t}
     \la[\delta \hat U(t),\delta \hat U(0)]\ra_{\D}$ and $J_{\B}(\w)\equiv \frac{1}{2}\int^{\infty}_{-\infty}\!\d t\,
  e^{i\w t}
     \la[\delta \hat V(t),\delta \hat V(0)]\ra_{\B}$, as
\cite{Wei08,Kle09,Yan05187}
\begin{align}\label{J1}
J_{\D}(\w)=\frac{2\lambda\gamma \w}{\w^2+\gamma^2}\ \ \ \  \text{and}\ \ \ \ 
  J_{\B}(\w)   
     =\frac{2\lambda'\w_0\zeta \w}{(\w^2-\w_0^2)^2+\w^2\zeta^2}.
\end{align}

\begin{figure}[h]
  \centering
  \includegraphics[width=1\columnwidth]{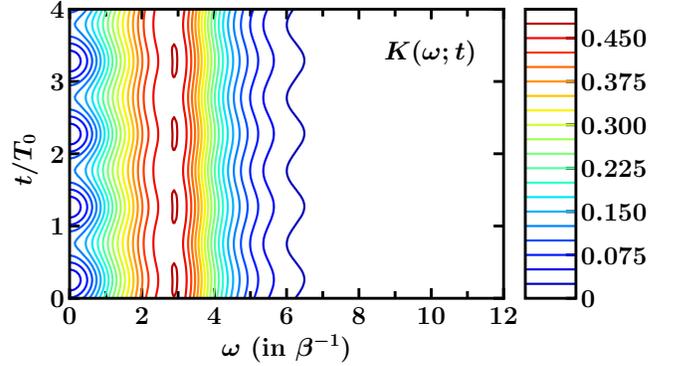}
  \caption{An example of $K(\w;t)$, with the same parameters used in \Fig{fig2}.}\label{fig3}
\end{figure}

To further explicitly exhibit the underlying non-Markovianity, we represent the kernel in the frequency domain as
\begin{align}
  K(\w;t)=\int_{0}^{\infty}\!\!{\rm d}\tau\, k(\tau;t)\cos(\omega\tau).
\end{align}
Note the periodicity in $t$ remains, i.e.
$
  K(\w;t) = K(\w;t+T_0).
$
 The \Fig{fig3} depicts the time--dependent frequency--resolved rate kernel, where this periodicity is manifest.
Therefore, we may do the Fourier expansion with respect to $t$,
\begin{subequations}
  \begin{align}
    K(\w;t)=\sum_{n=-\infty}^\infty K_n(\w)e^{-in\Omega t},
  \end{align}
 obtaining its components $\{K_n(\w)\}$ being
  \begin{align}
    K_n(\w)=\frac{1}{T_0}\int_{0}^{T_0}\!{\rm d}t\,K(\w;t)e^{in\Omega t}.
  \end{align}
\end{subequations}

\begin{figure}
  \includegraphics[width=1\columnwidth]{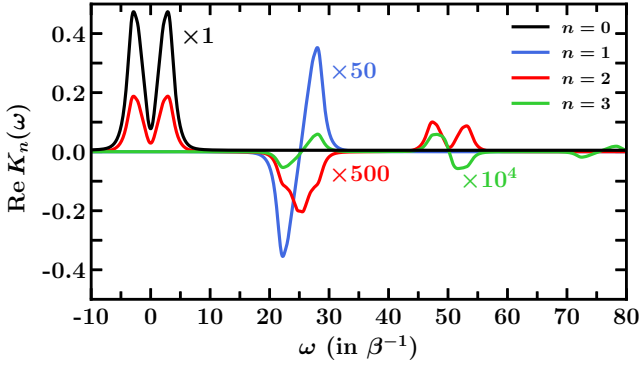}
    \caption{Real parts of the Fourier components $\{K_{n}(\w)\}$ of the rate kernels in unit of $\beta^{-1}$, with the same parameters used in \Fig{fig2}.
  }
  \label{fig4}
  \end{figure}
In \Fig{fig4}, we plot the real parts of the Fourier components $\{K_{n}(\w)\}$, exemplified with the same case as in \Fig{fig2}. These frequency--domain components exhibit the interplay between the external frequency, $\Omega$, and the so--called Floquet frequency, $\omega_{\F}$. The $\omega_{\F}$ is identified as the characteristic frequency of the Floquet Hamiltonian, $H_{\F}$, defined via 
$e^{-iH_{\F}T_0} \equiv \exp_+\big[-i\int_{0}^{T_0}\!\ud t\, H_{\T}(t)\big]$. As shown in the figure, in the case of $\beta\Omega=8\pi$ and $\beta\omega_{\F}\sim 3$, the peaks of $\{K_n(\omega)\}$ are centered at the integer multiples of $\Omega$. Meanwhile, each peak is split or deformed with split width being $2\omega_{\F}$. In SM, we give the analytical explanation of this phenomenon with the help of the Floquet theorem.

% From \Fig{fig4}, it is observed that the Fourier components are characterized by two important frequencies, the frequency of external field $\Omega$ and the Floquet frequency of the total Hamiltonian $\omega_{\F}$. Each peak in \Fig{fig4} appears around $m\Omega\pm\omega_{\F}$, with $m$ being an integer.  

\begin{figure}[h]
  \includegraphics[width=1\columnwidth]{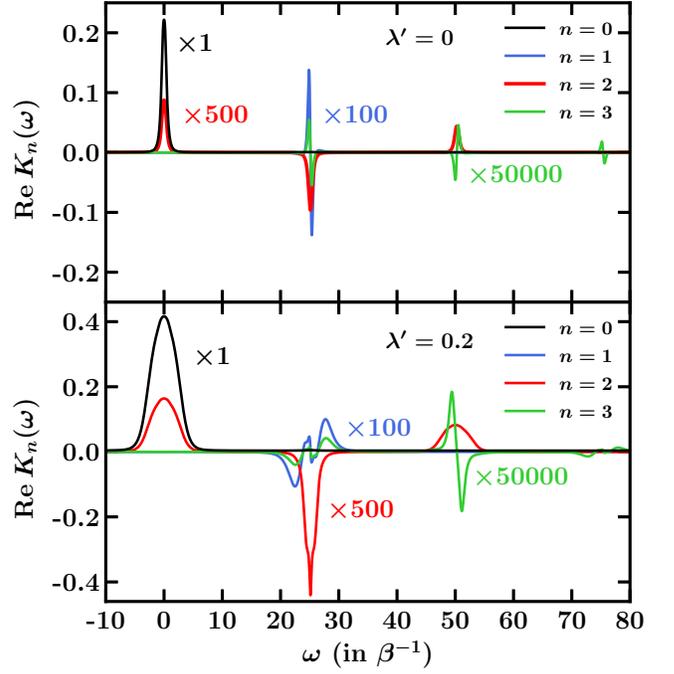}
    \caption{Real parts of the Fourier components evaluated with or without bridge fluctuation, in unit of $\beta^{-1}$. We set $\mathcal{E} = 0.2$, $\Omega=8\pi$, $E^{\circ}=-0.2$, $\la V\ra_{\B}=0.2$, $\lambda=0.2$, $\gamma=\omega_0=\zeta=1$ for both the cases, while $\lambda'=0$ and $0.2$ in the upper and lower panel, respectively.
  }
  \label{fig5}
  \end{figure}

\paragraph*{Herzberg--Teller--like mechanism.}
To conclude this Letter, we remark on the Herzberg--Teller--like coupling induced by the bridge fluctuation subject to effective modulation. 
In the absorption spectroscopy, the Herzberg--Teller mechanism manifests the non-Condon vibronic couplings, where the transition dipole moment involves the nuclear coordinate dependence. 
There exists a similar mechanism in our setting of ET with fluctuating bridge.
To illustrate this point, we first evaluate the Fourier components with or without bridge fluctuations. 
As shown in \Fig{fig5}, it is evident that the existence of fluctuating bridge would broaden, strengthen and deform each peak of the Fourier components.
This reflects the Herzberg--Teller--like mechanism, as analysed below.
In the high--frequency limit $\varphi(t) \ll 1$, the Hamiltonian in \Eq{HSB2} can be written in the form of
\begin{align}\label{27}
  H_{\T}'(t) =& H_{\tS}^{0}+h_{\E}- |{\rm A}\ra\la {\rm A}|\delta\hat U - \hat Q\delta\hat V
\nl &  
  -\hat \mu^{-}_{\T}\ti E^{(+)}(t)-\hat \mu^{+}_{\T}\ti E^{(-)}(t)
\end{align}
with
$
H_{\tS}^{0}=(E^{\circ}+\lambda)|{\rm A}\ra\la {\rm A}|+\la V\ra_{\B}|({\rm D}\ra\la {\rm A}|+|{\rm A}\ra\la {\rm D}|),
$ 
$\hat \mu_{\T}^{+}=|{\rm A}\ra\la {\rm D}|(\la V\ra_{\B}-\delta\hat V)=(\hat \mu_{\T}^{-})^{\dg}$ and $\ti E^{+}(t)=i\varphi(t)=[\ti E^{-}(t)]^{\dg}$. The last two terms in \Eq{27} can be seen as an effective dipole--field coupling, where the total dipole involves the bridge degrees of freedom. This resembles a type of Herzberg--Teller coupling. \cite{Zha16204109}
In the Markovian and high--frequency limits, the rate constant is given by
\begin{align}\label{DEOM-k}
  K'_0 = K_0(\omega = 0), 
\end{align}
which shall be compared with the perturbative  result in \Eq{k_ref}.
\begin{figure}
  \includegraphics[width=1\columnwidth]{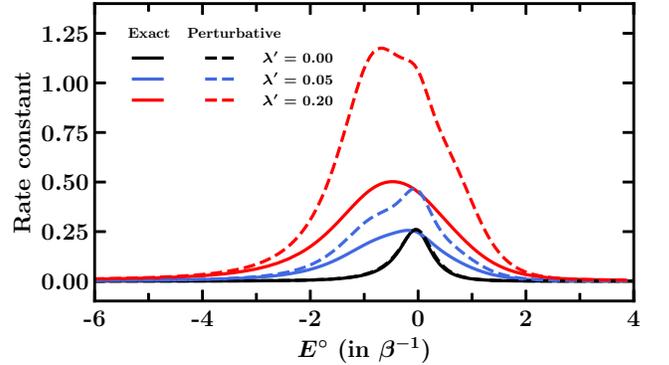}
    \caption{Rate constants $K_0$ and $K_0'$ versus $E^{\circ}$ in unit of $\beta^{-1}$, with different strengths of bridge fluctuation, $\lambda'$. Other parameters are the same with that in \Fig{fig5}.
  }
  \label{fig6}
  \end{figure}
In \Fig{fig6}, we plot the rate constants $K_0$ and $K_0'$ versus $E^{\circ}$, with different strengths of bridge fluctuation.
As shown in \Fig{fig6}, in the regime of $E^{\circ}+\lambda \sim 0$, the perturbative results with larger $\lambda'$  depart more from the nonperturbative results. 
 
\paragraph*{Summary.} In summary, we study the ET under the Floquet modulation occurring in the DBA systems. 
  The rate kernels are constructed and evaluated via the exact projected disspaton--equation--of--motion formalism.
  This enables us to investigate the interplay between the intrinsic non-Markovianity and driving periodicity. 
The bridge fluctuations manifestly affect the rate kernel, exhibiting a Herzberg--Teller--like mechanism subject to effective modulation.
 It is anticipated that our study will benefit the design of ET manipulation in molecular systems. 
 The same method can be also applied to such as excitation energy transfer in light harvest systems. \cite{Tha181243}
 
%  on the dynamics of electron transfer in the strongly correlated molecular systems. 
%
 
%Further, the exact constructed rate kernel provides a comprehensive approach to investigating the dynamics in no matter the ET or excitation energy transfer under external driving fields. 

\begin{acknowledgements}
  The authors thank the support from the Ministry of Science and Technology of China (Grant No.\ 2021YFA1200103) and the National Natural Science Foundation of China (Grant Nos.\
  22103073, 22173088 and 21903078).
  YS and HJZ thank also the
  partial support from the College Students' Innovative Entrepreneurial Training Plan Program (2020).
  YW and ZHC thank also the
  partial support from GHfund B (20210702).
\end{acknowledgements}

\providecommand{\latin}[1]{#1}
\makeatletter
\providecommand{\doi}
  {\begingroup\let\do\@makeother\dospecials
  \catcode`\{=1 \catcode`\}=2 \doi@aux}
\providecommand{\doi@aux}[1]{\endgroup\texttt{#1}}
\makeatother
\providecommand*\mcitethebibliography{\thebibliography}
\csname @ifundefined\endcsname{endmcitethebibliography}
  {\let\endmcitethebibliography\endthebibliography}{}

\end{document}

\title{Supplementary material on ``Electron transfer under the Floquet modulation in donor--bridge--acceptor systems''
}
%%%
\author{Yu Su}
\author{Zi-Hao Chen}
\author{Haojie Zhu}
\author{Yao Wang} 
\email{wy2010@ustc.edu.cn}
\author{Lu Han}
\author{Rui-Xue Xu}
%\email{rxxu@ustc.edu.cn}
\author{YiJing Yan}
\email{yanyj@ustc.edu.cn}
\affiliation{%Hefei National Laboratory for Physical Sciences at the Microscale and 
Department of Chemical Physics,
University of Science and Technology of China, Hefei, Anhui 230026, China}

\date{\today}

\begin{abstract}
This supplementary material contains: (i) a brief introduction to DEOM, (ii) the rigours proof of the periodicity   of rate kernels, and (iii) the analytical analysis of the frequency--resolved rate kernels with the help of the Floquet theorem.   
\end{abstract}

\maketitle

\section{A brief introduction to DEOM}

 Consider an arbitrary system  coupled to the environment with the temperature $T$.
Generally, the composite Hamiltonian can be decomposed into the system, the system--environment coupling and the environment parts as
\be\label{Hall}
H_{\T}(t)=H_{\s}(t)
+\sum_a\hat Q_a(t)\hat F_{a}+h_{\B}.
\ee
Here,  the dissipative modes $\{\hat Q_a\}$ are rather arbitrary,
while the environment Hamiltonian are adopted to be Gaussian.
This requires not only $h_{\B}$ be harmonic but also
the hybrid reservoir modes $\{\hat F_a\}$ be linear.
That is
\be\label{gwmodel}
h_{\B} =\frac{1}{2}\sum_{j}\omega_{j}(p_{j}^2+x_{ j}^2) \ \ \ \ {\rm and}\  \ \ \ 
\hat F_{a}=\sum_j c_{aj}x_{j}.
\ee
Here, the oscillators of frequency $\{\w_{j}\}$, with position $\{x_{j}\}$ and momentum $\{p_{j}\}$, are  coupled to the system with strength $\{c_{a j}\}$. 
As in the main text, we set $\hbar=1$ and $\beta=1/(k_BT)$,
with $k_B$ being the Boltzmann constant. Specifically, \Eq{Hall} with \Eq{gwmodel} covers both the Hamiltonians in  Eqs.(11) and (12) of the main text.

For the Gaussian environment (also named as ``bath'' ), its influences on the system is totally characterized by the
spectral density,  $J_{ab}(\w)=\frac{\pi}{2}\sum_jc_{a j}c_{b j}\delta(\w-\w_{j})$ for $\w\geq 0$, and it is related to the statistical bath correlation via the fluctuation--dissipation theorem as \cite{Wei08,Yan05187} ,
\begin{align}\label{FBt_corr}
\la \hat F_{a}(t)\hat F_{b}(0)\ra_{\B}&=
\frac{1}{\pi}\int^{\infty}_{-\infty}\!\!
\d\w \frac{e^{-i\w t} J_{ab}(\w)}{1-e^{-\beta\w}}
%\nl&
\simeq \sum_{k=1}^K\eta_{abk}e^{-\gamma_{k} t}.
\end{align}
Here, $\hat F_{a}(t)\equiv e^{ih_{\B}t}\hat F_{a}e^{-ih_{\B}t}$
is the hybridizing bath operator in the $h_{\B}$-interaction picture.
%%%
The average $\la\hat O\ra_{\B}\equiv {\rm tr}_{\B}(\hat Oe^{-\beta h_{\B}})/
{\rm tr}_{\B}(e^{-\beta h_{\B}})$ runs over
the bare--bath thermal equilibrium ensembles for any operator $\hat O$.
The multi-exponential decomposition in the the last identity
can be readily achieved with some advancing sum-over-pole schemes.
 \cite{Hu10101106,Hu11244106}
The exponents from the poles of Bose function and the corresponding pre-exponents coefficients are both real.
On the other hand, those originating from the spectral density are either real or complex conjugate paired. Therefore, we can define $\bar k$ such that $\gamma_{\bar k}=\gamma_{k}^{\ast}$.

Dynamical variables in DEOM are 
the dissipaton--augmented--reduced density operators
(DDOs):\cite{Yan14054105,Yan16110306,Zha18780}
$
  \rho^{(n)}_{\bf n}(t)
\equiv {\rm tr}_{\B}\Big[
  \Big(\prod_{ak} \hat f_{ak}^{n_{ak}}\Big)^{\circ}\rho_{\T}(t)
 \Big]
$.
%%%
Here, ${\bf n}\equiv \{n_{ak}\}$ 
specifies the configuration of the total $n$--dissipatons,
where  $n=\sum_{ak}n_{ak}$, with $n_{ak}\geq 0$
for bosonic dissipatons.
%%%
The reduced system density operator
is just $\rho_{\bf 0}^{(0)}(t)$.
The DEOM reads \cite{Yan14054105}
\begin{align}\label{DEOM}
 \dot\rho^{(n)}_{\bf n}=
& -[H_{\tS}(t),\rho^{(n)}_{\bf n}]-\sum_{ak} n_{ak} \gamma_{k}\rho^{(n)}_{\bf n}
  -i\sum_{ak}\Big[\hat Q_{a}(t),\rho^{(n+1)}_{{\bf n}_{ak}^+}\Big]
%\nl&
  -i\sum_{abk}n_{ak}\Big[\eta_{abk} \hat Q_{a}(t)\rho^{(n-1)}_{{\bf n}_{bk}^-}-\eta_{ab\bar k}^{\ast}
  \rho^{(n-1)}_{{\bf n}_{bk}^-}\hat Q_{a}(t)\Big].
\end{align}
This describes the same dynamics of the HEOM formalism.\cite{Tan89101,Tan906676,Tan06082001,Yan04216,Xu05041103,Xu07031107}
It is worth emphasizing that the DEOM theory contains not only the above hierarchical dynamics, but also the dissipaton algebra. \cite{Wan20041102}
%%%
For brevity, we may express the \Eq{DEOM} in the form of
\be\label{sdeom}
\dot{\bm\rho}(t)=-i\bm{\mathcal{L}}(t){\bm\rho}(t).
\ee
This resembles $\dot{\rho}_{\T}=-i{\cal L}_{\T}(t)\rho_{\T}$ with mapping the total system--plus--bath composite  Liouvillian to  the DEOM--space dynamics generator as defined in
\Eq{DEOM}, ${\cal L}_{\T}(t)\rightarrow \bm{\mathcal{L}}(t)$,
and
$
\rho_{\T}(t)\rightarrow {\bm \rho}(t)=\{\rho_{\bf n}^{(n)}(t); n=0,1,2,\cdots\}.
$
Equation (\ref{sdeom}) is exactly Eq.(13) of the main text.

\section{Proof of the periodicity   of rate kernels}
We first review the projected DEOM approach to rate kernels in more details [cf.\,Eqs.(13)-(17) of the main text].
 From the DEOM (\ref{sdeom}), we know that the propagator
\be
  \bm{{\cal U}}(t,\tau)=\exp_+\Big[-i\int_{\tau}^{t}\d\tau'\,\bm{{\cal L}}(\tau')\Big]
\ee
governs the dynamics of the DEOM state function $\bm{\rho}\equiv\{\rho_{\bf n}^{(n)}\}$.
To proceed, define DEOM--space projection operators, $\bm{{\cal P}}$ and $\bm{{\cal Q}}=\bm{{\cal I}}-\bm{{\cal P}}$, for partitioning $\bm{\rho}\equiv\{\rho_{\bf n}^{(n)}\}$ into the population and coherent components,
respectively:\cite{Zha163241}
\be\label{HEOM_PQ_def}
\begin{split}
 \bm{{\cal P}}\bm{\rho}(t)&= \Big\{\sum_{a} \rho^{(0)}_{aa}(t)|a\ra\la a|;\ 0,\,0,\,\cdots\,\Big\} \equiv {\bm p}(t)\,,
\\
 \bm{{\cal Q}}\bm{\rho}(t)&= \Big\{\sum_{a\neq b} \rho^{(0)}_{ab}(t)|a\ra\la b|;\ \rho^{(n>0)}_{\bf n}(t)\Big\} \equiv {\bm\sigma}(t)\,.
\end{split}
\ee
We can now recast the DEOM (\ref{sdeom}) in terms of
\be\label{sep}
\begin{bmatrix}
	\dot{\bm p}(t) \\ \dot{\bm\sigma}(t)
\end{bmatrix}
=-i \begin{bmatrix}
 	\bm{{\cal P}{\cal L}}(t)\bm{{\cal P}}  & \bm{{\cal P}{\cal L}}(t)\bm{{\cal Q}}\\
	\bm{{\cal Q}{\cal L}}(t)\bm{{\cal P}}  & \bm{{\cal Q}{\cal L}}(t)\bm{{\cal Q}}
\end{bmatrix}
\begin{bmatrix}
	{\bm p}(t) \\ {\bm\sigma}(t)
\end{bmatrix}.
\ee
The formal solution to ${\bm\sigma}(t)$ reads
%\be\label{coht}
% {\bm\sigma}(t)=-i\int_{0}^{t}\!d\tau\,\bm{\bar{{\cal G}}}(t,\tau)\bm{{\cal Q}{\cal L}}(\tau)%%{\bm p}(\tau)
%  +\bm{\bar{{\cal G}}}(t,0){\bm\sigma}(0),
%\ee
\be\label{coht}
 {\bm\sigma}(t)=-i\int_{0}^{t}\!\d\tau\,\bm{{\cal Q}}\bm{{{\cal U}}}(t,\tau)\bm{{\cal Q}{\cal L}}(\tau){\bm p}(\tau).
\ee
Equation (\ref{coht}) contains no inhomogeneous term, as we assume initially the total density operator is factorized, and thus ${\bm\sigma}(0)=0$.
Noting that $
 -i\bm{{\cal PL}}(t){\bm p}(t) = 0$,\cite{Zha163241} the first identity of \Eq{sep} then has the expression
\be\label{popt}
  \dot{\bm p}(t)= \int_{0}^{t}\!\d\tau\,\bm{\tilde{K}}(t-\tau;t){\bm p}(\tau),
\ee
with the rate kernel being formally of
\be\label{calK_formal}
 \bm{\tilde{K}}(t-\tau;t) = -\bm{{\cal PL}}(t)\bm{{\cal Q}}
 \bm{{\cal U}}(t,\tau)
\bm{{\cal QL}}(\tau)\bm{{\cal P}}.
\ee
Apparently, $-k(t-\tau;t)$ and $k'(t-\tau;t)$ in Eq. (1) of main text are the $|{\rm D}\ra\la {\rm D}| \rightarrow |{\rm D}\ra\la{\rm D} |$ and $|{\rm A}\ra\la {\rm A}| \rightarrow |{\rm D}\ra\la{\rm D} |$ components of $\bm{\tilde{K}}(t-\tau;t)$, respectively. That is to say,
\begin{align}
-k(t-\tau;t)&=\big\la\!\big\la  |{\rm D}\ra\la{\rm D} |, 0,0, \cdots \big|\big| \bm{\tilde{K}}(t-\tau;t)\big|\big|\,|{\rm D}\ra\la{\rm D} |, 0,0, \cdots\big\ra\!\big\ra ,
\\
k'(t-\tau;t)&=\big\la\!\big\la  |{\rm D}\ra\la{\rm D} |, 0,0, \cdots \big|\big| \bm{\tilde{K}}(t-\tau;t)\big|\big|\,|{\rm A}\ra\la{\rm A} |, 0,0, \cdots\big\ra\!\big\ra.
\end{align}

Now we can  prove that the kernel in \Eq{calK_formal}  satisfies
\be 
 \bm{\tilde{K}}(t-\tau;t+T_0)=\bm{\tilde{K}}(t-\tau;t).
\ee
To this end, we note that 
$\bm{{\cal L}}(t+T_0)=\bm{{\cal L}}(t)$ according to our Floquet setting. This leads to
\be
\bm{\tilde{K}}(t-\tau;t+T_0)\equiv -\bm{{\cal PL}}(t+T_0)\bm{{\cal Q}}
 \bm{{\cal U}}(t+T_0,\tau+T_0)
\bm{{\cal QL}}(\tau+T_0)\bm{{\cal P}}=\bm{{\cal PL}}(t)\bm{{\cal Q}}
 \bm{{\cal U}}(t,\tau)
\bm{{\cal QL}}(\tau)\bm{{\cal P}}=\bm{\tilde{K}}(t-\tau;t).
\ee
Here we have used 
\begin{align}
\bm{{\cal U}}(t+T_0,\tau+T_0)&=\exp_{+}\Big[-i\int_{\tau+T_0}^{t+T_0}\d\tau'\,\bm{{\cal L}}(\tau')\Big]=\exp_{+}\Big[-i\int_{\tau}^{t}\d\tau'\,\bm{{\cal L}}(\tau'+T_0)\Big]
\nl & 
=\exp_{+}\Big[-i\int_{\tau}^{t}\d\tau'\,\bm{{\cal L}}(\tau')\Big]=\bm{{\cal U}}(t,\tau).
\end{align}
Therefore, we conclude $k(t-\tau;t)$ and $k'(t-\tau;t)$ is periodic with respect to their second parameters.

\section{Analysis of the frequency--resolved rate kernels}
To proceed, we first recast \Eq{calK_formal} as
\be 
 \bm{\tilde{K}}(\tau;t) = -\bm{{\cal PL}}(t)\bm{{\cal Q}}
 \bm{{\cal U}}(t, t-\tau)
\bm{{\cal QL}}(t-\tau)\bm{{\cal P}}.
\ee
Then, according to the Floquet theorem, the total propagator can be decomposed as \cite{Res16250401}
\begin{align}
U_{\T}(t,0)\equiv T_{+}e^{-i\int_{0}^{t}\!{\rm d}\tau\,H_{\T}(\tau)}
%\nl &
=e^{-iK_{\T}^{\F}(t)}e^{-iH_{\T}^{\F}t},
\end{align}
with the Floquet Hamiltonian $H_{\T}^{\F}$ defined as
\be 
e^{-iH_{\T}^{\F}T_0}\equiv U(T_0,0)
\ee
and the stroboscopic kick operator $K_{\T}^{\F}(t)$
\be 
e^{-iK_{\T}^{\F}(t)}\equiv U(t,0)e^{iH_{\T}^{\F}t}.
\ee
The  stroboscopic kick operator satisfies $K_{\T}^{\F}(0)=K_{\T}^{\F}(nT_0)=0$.
Then, in the DEOM--space we may obtain
\be 
 \bm{\tilde{K}}(\tau;t) = -\bm{{\cal PL}}(t)\bm{\mathcal{Q}}
 \bm{{\cal U}}_{\K}(t)\bm{{\cal G}}_{\F}(\tau)\bm{{\cal U}}^\dagger_{\K}(t-\tau)
\bm{{\cal QL}}(t-\tau)\bm{{\cal P}},
\ee
with $\bm{{\cal G}}_{\F}(\tau)$, $\bm{{\cal U}}_{\K}(\tau)$ and $\bm{{\cal U}}^\dagger_{\K}(t-\tau)$ are the $\exp\{-i\tau [H_{\T}^{\F}, \,\cdot\,]\}$, $\exp\{-i[K_{\T}^{\F}(\tau),\,\cdot\,]\}$ and $\exp\{i[K_{\T}^{\F}(t-\tau),\,\cdot\,]\}$  mapped to DEOM space, respectively.

% ,  followed by the projection $\bm{{\cal Q}}$. 

 Since the system is driven periodically, we can expand the rate kernel as 
\begin{align}
  {\bm{\bar{K}}(\tau;t)} = \sum_{m=-\infty}^\infty\sum_{n=-\infty}^\infty {\bm{\mathcal{A}}}_m {\bm{\mathcal{G}}}_{\F}(\tau){\bm{\mathcal{B}}}_n e^{-i(m+n)\Omega t}e^{in\Omega \tau},
\end{align}
where 
\begin{align}
  {\bm{\mathcal{A}}}_m \equiv -\frac{i}{T_0}\int_0^{T_0}\!\ud t\, \bm{{\cal PL}}(t)\bm{\mathcal{Q}}
  \bm{{\cal U}}_{\K}(t) e^{im\Omega t}
\end{align}
and 
\begin{align}
  {\bm{\mathcal{B}}}_n \equiv -\frac{i}{T_0}\int_0^{T_0}\!\ud t\, \bm{{\cal U}}^\dagger_{\K}(t)
  \bm{{\cal QL}}(t)\bm{{\cal P}}e^{in\Omega t}.
\end{align}
By changing the summation indices, the rate kernel can be recast as 
\begin{align}
  {\bm{\bar{K}}(\tau;t)} = \sum_{m=-\infty}^\infty e^{-im\Omega t}\sum_{n=-\infty}^{\infty} \bm{\mathcal{A}}_{m-n}\bm{\mathcal{G}}_{\F}(\tau)\bm{\mathcal{B}}_n e^{in\Omega\tau}.
\end{align}
Therefore, the corresponding Fourier components in frequency domain are given by 
\begin{align}
  \bm{\bar{K}}_m(\omega) = \int_{0}^\infty\!\!\ud\tau\sum_{n=-\infty}^\infty \bm{\mathcal{A}}_{m-n}\bm{\mathcal{G}}_{\F}(\tau)\bm{\mathcal{B}}_n e^{in\Omega\tau}\cos(\omega\tau).
\end{align}
Since the Floquet Hamiltonian is independent of time, one could express the Fourier components formally via the eigenstates of $\bm{\mathcal{L}}_{\F}$, the mapping of $[H_{\T}^{\F},\,\cdot\,]$ in DEOM--space, as 
\begin{align}
  \bm{\bar{K}}_m(\omega) = \int_{0}^\infty\!\!\ud\tau\sum_{n=-\infty}^\infty \sum_{\omega_{\F}}\bm{\mathcal{A}}_{m-n}(\omega_{\F})\bm{\mathcal{B}}_n(\omega_{\F}) e^{i(n\Omega-\omega_{\F})\tau}\cos(\omega\tau). 
\end{align}
That is to say, we set
$
\bm{\mathcal{B}}_n=\sum_{\w_{\F}}\bm{\mathcal{B}}_n(\w_{\F}), 
$
such that 
$
\bm{\mathcal{A}}_{m-n}\bm{\mathcal{L}}_{\F}\bm{\mathcal{B}}_n(\w_{\F})=\w_{\F}\bm{\mathcal{A}}_{m-n}(\w_{\F})\bm{\mathcal{B}}_n(\w_{\F}).
$
The integral over $\tau$ gives the basic profiles of the peak (real part) of $\bm{\bar{K}}_m(\omega)$ proportional to
%\begin{align}\label{29}
%  \frac{1}{2i}\bigg[ \frac{1}{\omega-(n\Omega-\omega_{\F})} - \frac{1}{\omega+(n\Omega-%\omega_{\F})}\bigg]+ 
%  \frac{\pi}{2}\big[ %\delta(\omega-n\Omega+\omega_{\F})+\delta(\omega+n\Omega-\omega_{\F}) \big].
%\end{align}
\begin{align}\label{29}
 \alpha\Big[\frac{1}{\omega-(n\Omega-\omega_{\F})} - \frac{1}{\omega+(n\Omega-\omega_{\F})}\Big]+\alpha'\Big[\frac{1}{\omega-(n\Omega+\omega_{\F})} - \frac{1}{\omega+(n\Omega+\omega_{\F})}\Big],
\end{align}
with $\alpha$ and $\alpha'$ being the coefficients.
Equation (29) results in the peak--split phenomenon of $\bm{\bar{K}}_m(\omega)$.

% Since the Floquet Hamiltonian is independent of time, the corresponding propagator $\bm{\mathcal{G}}_{\F}(\tau)$ must take the behavior like $e^{-i\omega_{\F}\tau}$, with $\omega_{\F}$ being the characteristic frequency of Floquet Hamiltonian. This results in the peak--split phenomenon of $\bm{\bar{K}}_m(\omega)$.

%\bibliographystyle{./aiptit.bst}
%\bibliography{./bibrefs}